\documentclass[prb,showpacs,superscriptaddress,twocolumn,notitlepage]{revtex4-1}

\usepackage{amsmath}
\usepackage{amssymb}
\usepackage{amsfonts}
\usepackage{graphicx}
\usepackage{multirow}
\usepackage{chngcntr}
\usepackage{bm}
\usepackage[T1]{fontenc}

\usepackage{color}
\usepackage[usenames,dvipsnames]{xcolor}

\usepackage{natbib}
\usepackage[final=true]{hyperref}
\hypersetup{
    colorlinks=true,
    linkcolor=Maroon,
    filecolor=magenta,      
    urlcolor=cyan,
    citecolor=PineGreen,    
}

\newcommand{\be}{\begin{eqnarray}}
\newcommand{\ee}{\end{eqnarray}}

\begin{document}


\title{Coupled oscillators model for hybridized optical phonon modes in contacting nanosized particles and quantum dot molecules}


\author{S.~V.~Koniakhin$^*$}
\affiliation{Center for Theoretical Physics of Complex Systems, Institute for Basic Science (IBS), Daejeon 34126, Republic of Korea}
\email{kon@mail.ioffe.ru}

\author{O.~I.~Utesov}
\affiliation{Department of Physics, St. Petersburg State University, St.Petersburg 199034, Russia}
\affiliation{Petersburg Nuclear Physics Institute NRC ``Kurchatov Institute'', Gatchina 188300, Russia}
\affiliation{St. Petersburg School of Physics, Mathematics, and Computer Science, HSE University, St. Petersburg 190008, Russia}

\author{A.~G.~Yashenkin}
\affiliation{Petersburg Nuclear Physics Institute NRC ``Kurchatov Institute'', Gatchina 188300, Russia}
\affiliation{Department of Physics, St. Petersburg State University, St.Petersburg 199034, Russia}

\begin{abstract}
  
Modification of optical phonon spectra in contacting nanoparticles as compared to the single ones is studied. Optical phonons in dielectric and semiconducting particles obey the Euclidean metric Klein-Fock-Gordon equation with Dirichlet boundary conditions. The latter is supposed to be solved numerically for manifolds of interpenetrating spheres. It is proposed to replace this problem
with the simpler-to-solve coupled oscillators model (COM), where an oscillator is attributed to each phonon mode of a particle and the particles overlap leads to appearance of additional couplings for these oscillators with the magnitude proportional to the overlapped volume. For not too big overlaps this model describes solutions of the original eigenvalue problem on a good level of accuracy. In particular, it works beyond isotropic s modes, which has been demonstrated for p modes in dimer and also for tetramer. It is proposed to apply COM for the description of recently manufactured dimer nanoparticles and quantum dots. The obtained results are in agreement with the dynamical matrix method for optical phonons in nanodiamonds. The latter is used to demonstrate that the van der Waals contacts between faceted particles lead to very small modifications of the optical phonon spectra, which therefore could be neglected when discussing the propagation of vibrational excitations via a nanopowder. The possibility to distinguish between dimerized and size-distributed single particles from their Raman spectra is also considered.
\end{abstract}

\maketitle

\section{Introduction}

Raman spectrosopy is a powerful tool widely used for characterization of modern nanostructured materials including nanoparticles, nanorods, and two-dimensional nanostructures,  which gives a precise energy fingerprint of the excitations peculiar for the material like phonons \cite{gouadec2007raman,yadav2006low,arora2007raman,dzhagan2007resonant,dzhagan2009influence}, magnons\cite{dietz1971infrared,martin1977resonant,iliev2010phonon}, excitons\cite{zucker1987resonant,brewster2009exciton}, etc. Presently, the Raman spectra measurements are the standard characterization procedure for carbon materials\cite{surovtsev1999effect,surovtsev2017effect,vinogradov2018structure} and various applications-oriented nanomaterials \cite{ferrari2004raman,kudryavtsev2022raman,korepanov2017carbon,gao2019determination,sachkov2019localization,stehlik2015size,stehlik2016high,stehlik2021size,ekimov2022size}. Combining simplicity of implementation, nondestructive nature and versatile data obtained from spectra analysis, the Raman spectroscopy contributes significantly to modern nanotechnology and material science.

On the basic level, the structure of optical phonon lines obtained by Raman spectroscopy allows determination of the composition of a material. As far as nanoparticles are concerned, the more sophisticated theoretical approach applied to fit Raman can yield much more parameters, i.e., the nanoparticles size $L$, the deviation $\delta L$ in the size distribution function, the lattice impurities concentration and the geometrical shape (faceting)~\cite{ourBench}. The nanoparticles size can be roughly estimated from the phonon confinement model \cite{richter1981one,campbell1986effects,osswald2009phonon,korepanov2017carbon,paillard1999improved,faraci2006modified,zi1997comparison} or with higher accuracy (including the standard deviation in the size distribution) from the joint\cite{ourDMM} Dynamical Matrix Method \cite{born1955dynamical} -- Bond Polarization Model (DMM-BPM) theory \cite{snoke1993bond,cheng2002calculations}. The nanoparticle shape (faceting) also affects the main peak position and overall peak structure~\cite{ourEKFG} in Raman spectra of nanoparticles. Information about the type and the concentration of lattice impurities can be obtained from the broadening of the Raman peak for nanoparticles \cite{our3,our4,ourBench,utesov2021effects} and also for bulk materials \cite{hass1992lattice,spitzer1993isotopic,hanzawa1996disorder,surovtsev1999effect,poklonskaya2015raman}. This picture is actual for dielectric diamond nanoparticles as well as for crystalline Si, Ge, GaAs, CdTe and many other types of semiconductor quantum dots. 

Recently, such promising objects as quantum dot (QD) molecules have been synthesized and now they are a subject of extensive ongoing research \cite{panfil2019electronic,cui2019colloidal,cui2021neck,cui2021semiconductor}. QD dimers are investigated to find applications for biomolecules sensing and for nanoantennas with controlled polarization. An interesting feature that QD dimers demonstrate is their electronic structure matching with the one of molecules. Significant progress has been obtained in fine control of the neck thickness, which gives a possibility to tune precisely the coupling constant between the monomers.

At the same time, the hybridization of optical phonons in nanoparticles laying in contact with each other and/or in QD dimers can affect the Raman spectrum of an ensemble and, in particular, the crystalline Raman peak shift and its shape. These effects are evidently important for characterization of the whole ensemble mentioned above.

Also, complete understanding of the optical phonons hybridization and of the coupling in QD molecules opens up a possibility to theoretically describe the role of inter-particle contacts in the Raman spectra of nanopowders, tight nanoparticle agglomerates\cite{dideikin2017rehybridization}, strongly coupled QD Nanocrystal Solids \cite{zhao2021enhanced} and porous materials \cite{yang1994study,alfaro2008theory,alfaro2011raman,kosovic2014phonon,valtchev2013porous}. Currently, the Raman spectra analysis relies on the single particle properties only.

The present study is addressed to the problem of optical phonon modes in contacting and/or cojoined nanoparticles and quantum dots. We argue that along with direct calculation of these modes within the proposed geometry the simple coupled oscillators model (COM) could be used to qualitatively describe the corresponding vibrational spectra. After some insignificant improvements COM is sufficient to reproduce  the principal features of optical phonons hybridization. This statement is verified using both continuous scalar Euclidean Klein-Fock-Gordon model (EKFG, see Ref.~\cite{ourEKFG}) and atomistic DMM approaches. In strongly coupled regime of cojoined particles, the effect of hybridization on the Raman spectra is found to be pronounced, whereas in the case of weakly coupled (say, via the Van der Waals forces) particles the corresponding optical phonon frequency shifts are  negligible. The latter means that the theory of propagation of vibrational modes through a nanopowder can safely ignore the perturbations of phonon spectra in individual particles stemming from their contacts, operating only with the single particle spectral characteristics.

The rest of the paper is organized as follows. In Section II we start with description of the model of coupled oscillators to familiarize readers with the approach to be developed. Then we utilize EKFG for insight into the problem of optical phonons in the cojoined spherical particles. We demonstrate the similarity of these problems and build up consistent perturbation theory formulating COM for eigenvalues and eigenfunctions of intersecting spheres. In section III we use microscopic DMM model and compare its predictions with the yield of EKFG and coupled oscillators approaches. We also use DMM-BPM approach to study the case of two faceted particles contacting through weak van der Waals interaction. The last Section IV contains discussion of our results and main conclusions.

\section{Semi-quantitative approach: coupled oscillators and EKFG theory}

\begin{figure}
  \centering
  \includegraphics[width=6cm]{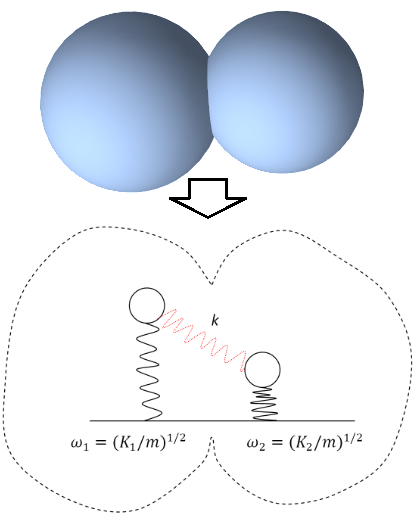}\\
  \caption{The eigenvalue problem of the Laplace operator \mbox{$\Delta \psi + q^2 \psi =0 $} with Dirichlet boundary conditions $\psi|_{\partial \Omega}=0$ on the manifold of two cojoined spheres has the solutions close to ones of two coupled harmonic oscillators model (COM).}\label{figspheres}
\end{figure}

Below we shall show that there is a good correspondence between the eigenmodes of EKFG approach for cojoioned particles and the classical problem of two coupled oscillators [coupled oscillators model (COM), see Fig.~\ref{figspheres}].

\subsection{Coupled oscillators}

The Lagrange function of two coupled oscillators reads
\be
  \mathcal{L} = \frac{m \dot{x}^2_1}{2} + \frac{m \dot{x}^2_2}{2} - \frac{k_1 x^2_1}{2} - \frac{k_2 x^2_2}{2} - \frac{k_{int} (x_1-x_2)^2}{2},
\ee
where two oscillators with frequencies $\omega^2_{1,2} = k_{1,2}/m$ are  assumed to be coupled by the spring with rigidity $k_{int}$. Newtonian equations of motion can be written as
\be
  m \ddot{x}_1 &=& - k_1 x_1 - k_{int}(x_1-x_2), \nonumber \\
  m \ddot{x}_2 &=& - k_2 x_2 - k_{int}(x_2-x_1).
\ee
To find the harmonic solutions, we rewrite them as
\be
\label{eq_newtonian_harmonic}
  ( k_1 + k_{int})x_1 - k_{int} x_2 &=& m \omega^2 x_1, \nonumber \\
  ( k_2 + k_{int})x_2 - k_{int} x_1 &=& m \omega^2 x_2.
\ee
The eigenvalues (squares of frequencies) are given by
\be \label{frcoup}
  \omega^2_{\pm} =
  \frac{k_1 + k_2 + 2 k_{int} \pm \sqrt{(k_1 - k_2)^2+ 4k^2_{int}}}{2m}.
\ee

In the particular resonant case $k_1=k_2=k$ one has
\be \label{eq_inresonance}
  \omega^2_{\pm} = \frac{k + 2 k_{int}}{m};  \frac{k}{m},
\ee
for $x_1 - x_2$ and $x_1 + x_2$ eigenfunctions respectively. We see that one resonant solution is not affected by the coupling, whereas the frequency of another one grows up linearly with $k_{int}$ parameter.

In the highly off-resonant case $|k_1-k_2| \gg k_{int}$ one gets
\be \label{eq_out_of_resonance}
  \omega^2_{\pm} &=& \frac{k_{1,2}+k_{int}}{m},
\ee
so the oscillators almost do not ``feel'' each other, but they feel the additional spring.

Notice that the generalization for $n>2$ coupled oscillators problem is straightforward.

\subsection{EKFG}

Continuous method of finding the eigenfunctions and the eigenvalues of the long-wavelength optical phonon modes applicable for {\it arbitrary} shape of the manifold $\Omega$ starts from evaluation of the Euclidean metric Klein-Fock-Gordon equation\cite{ourEKFG}:
\be \label{EKFG}
  ( \partial^2_t + C_1 \Delta +C_2) \psi =0, \quad \psi|_{\partial \Omega} =0,
\ee
where the second expression is referred to the Dirichlet boundary conditions. Parameters $C_{1,2}$ are related to the constants of the optical phonon dispersion in the long-wavelength limit via
\be
  \omega^2 &=& C_2 - C_1 q^2 \Longleftrightarrow \nonumber \\
  \label{eq_dispersion2}
  \omega(q) &\approx& \sqrt{C_2} - \frac{C_1}{\sqrt{C_2}} \frac{q^2}{2} \equiv \omega_0 - \alpha q^2,
\ee
$\omega_0$ is maximal optical mode frequency and $q^2$ is an eigenvalue of the corresponding boundary value problem
\be
  \Delta \psi + q^2 \psi = 0, \quad \psi|_{\partial \Omega} =0.
\ee
This equation can be solved numerically for arbitrary manifold using, e.g., Wolfram Mathematica. We dub $\psi$ a ``wave function'' below.


Unlike the yield of PCM, the realistic structure of the phonon spectrum in nanoparticle is discrete like in diamondoids \cite{jenkins1980raman,filik2006raman} and in fullerenes \cite{snoke1993bond}. The vibrational modes should resemble the standing wave-like eigenmodes of a resonator of the same shape as the nanoparticle or, more generally, electron orbitals in atom (see Refs.~\cite{ourDMM,ourEKFG}). Therefore, their classification into s, p, \textit{etc.} orbital-like classes makes perfect sense, at least  for spherical particles and their cojoined combinations. In particular, it means that the formation of the symmetric (``bonding'') and the anti-symmetric (``anti-bonding'') states in nanoparticle dimers (or quantum dot ``molecules'')  should occur on the same footing as it takes place in real atoms and molecules.

\begin{figure}
  \centering
  \includegraphics[width=8cm]{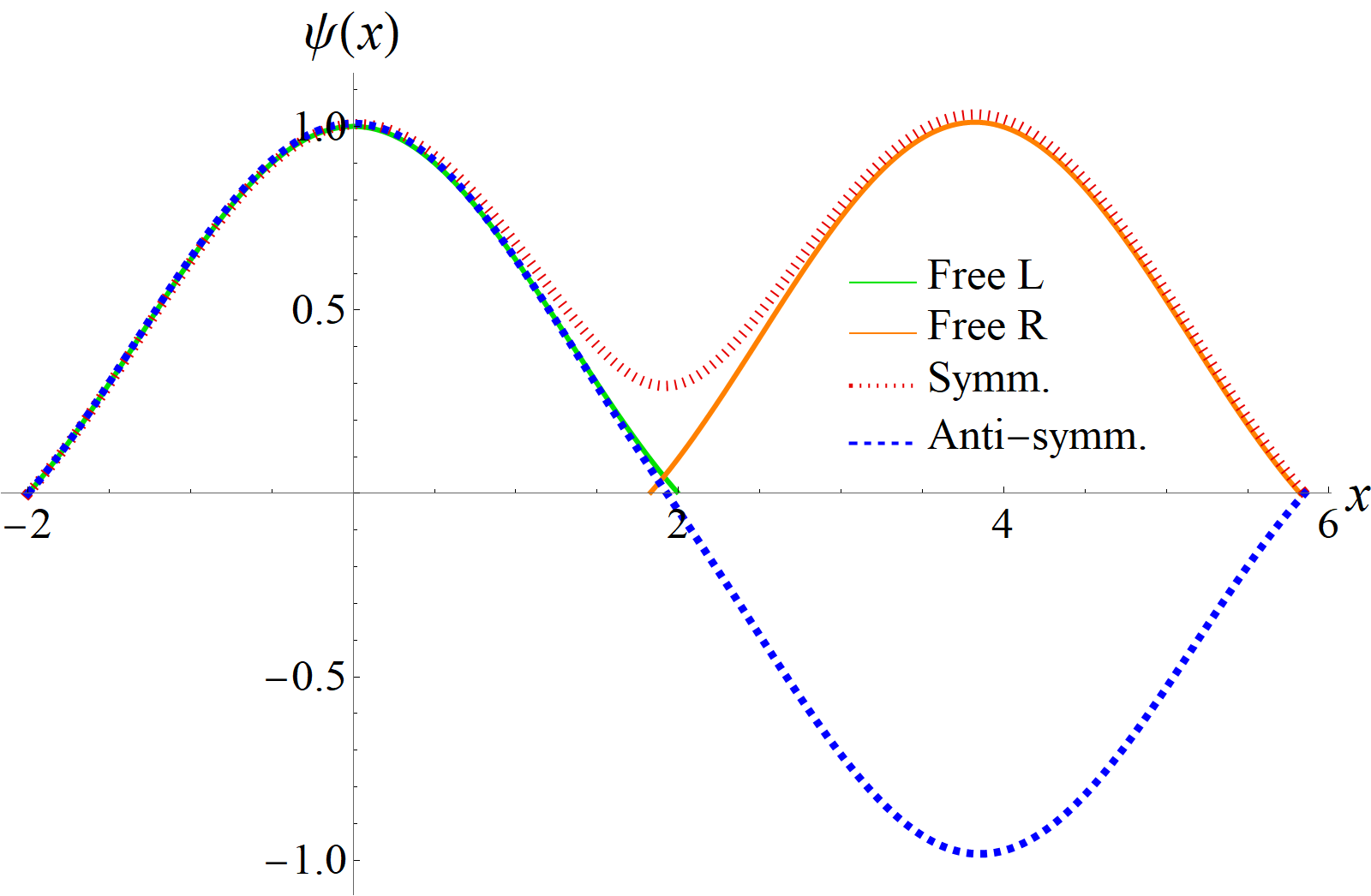}\\
  \caption{Phonon wave functions of s modes along \textit{x} axis (connecting the centers) for not contacting spheres with radii $R_1=R_2=2$ (green and orange, manually shifted left by $\delta r$, the eigenvalues $q_1^2=q_2^2=2.469$) and for symmetric and antisymmetric modes of the dimer with $\delta r = 0.2$ (red and blue, eigenvalues $q_1^2=2.447$, $q_1^2=2.471$). One sees that the anti-symmetric mode is very close to the mode of the free particles and has nearly the same eigenvalue. The eigenvalue of symmetric mode is strongly shifted and the wave function profile differs, in particular in the contact region. The symmetric mode in spheres corresponds to the antisymmetric mode in the coupled oscillators approach.}\label{fig_WFplots}
\end{figure}

Fig. \ref{fig_WFplots} shows optical phonon wave functions with smallest $q^2$ in two cojoined spheres  with radii \mbox{$R_1=R_2=2$} and the penetration length $\delta r = 0.2$.  It is important to underscore that the antisymmetric eigenfuction profile nearly coincides with the wave function profile of isolated sphere. The eigenvalue of antisymmetric mode is also very close to the one of isolated sphere. On the contrary, the symmetric wave function differs significantly from the isolated sphere wave function in the region of contact. The eigenvalue, corresponding to this function, is also downshifted with respect to the isolated sphere eigenvalue.

\begin{figure}
  \centering
  \includegraphics[width=8cm]{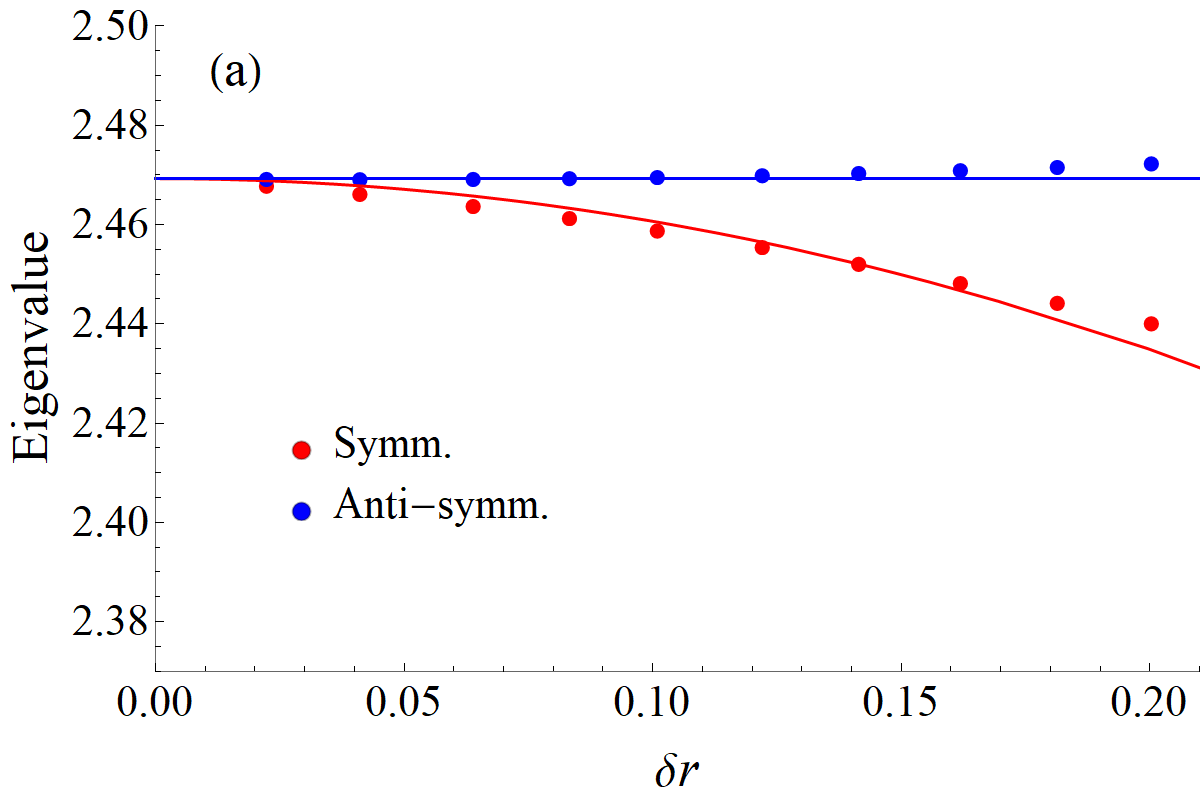}\\
  \includegraphics[width=8cm]{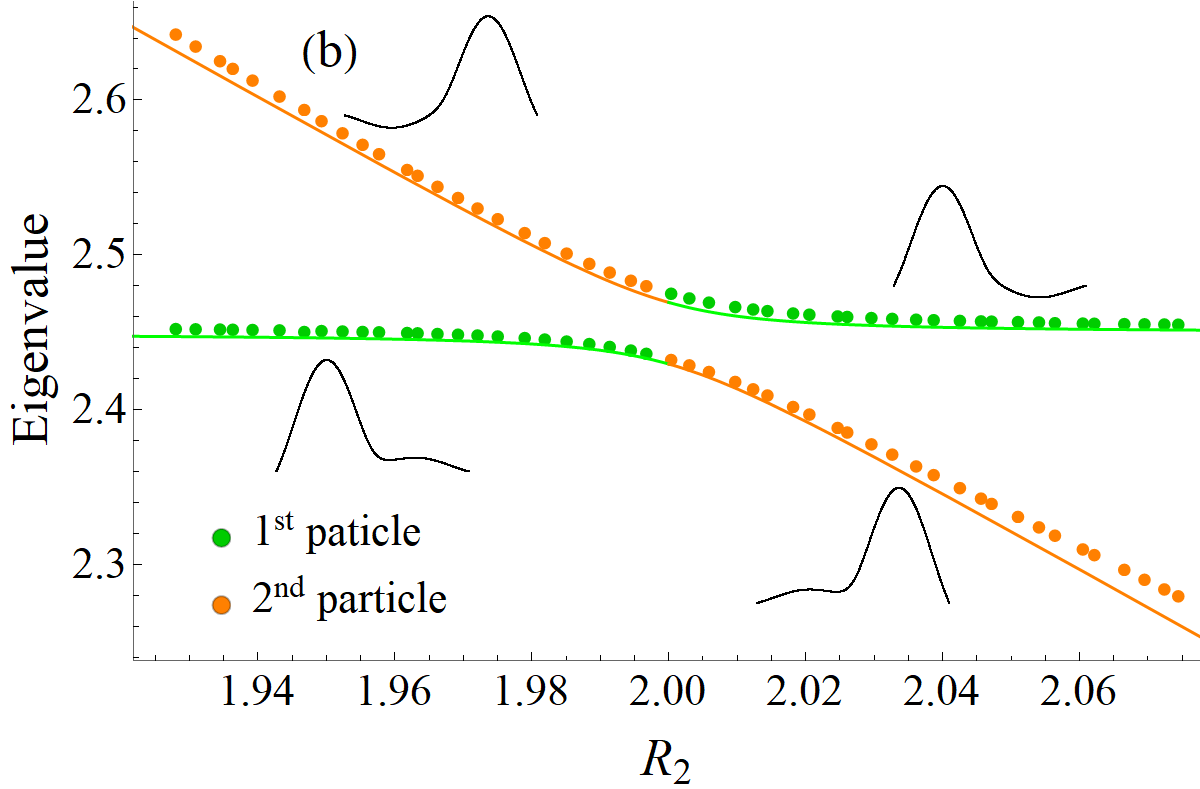}
  \caption{Panel (a). Two smallest eigenvalues of the Laplace operator in the system of two spheres of radii 2 as a function of penetration length $\delta r$ are given with red (symmetrical $\psi$) and blue (anti-symmetrical $\psi$) markers. Blue and red curves are the result of coupled oscillators approach, see Eq.~\eqref{frcoup}, for $k_1=k_2$. The dependence $k_{\rm int} \propto \delta r^2$ is used. Panel (b). Two upper Laplacian eingenvalues in the system of the two spheres with fixed penetration length $\delta r=0.21$ and the radius of left particle $R_1 = 2$ as a function of the right particle radius $R_2$. The green points correspond to the eigenfunctions concentrated in the first particle and the orange ones correspond to the eigenfunctions with domination inside the second particle. The solid colored curves represent the fit with the use of Eq.~\eqref{frcoup} for fixed $k_1$ and $k_{\rm int}$ and $k_2 = k_1 + {\rm const} \cdot (R_2-2)$. Freestanding black curves show sketched shapes of the wave functions.}\label{fig_modelcase}
\end{figure}

Fig.~\ref{fig_modelcase} shows two smallest eigenvalues for two interpenetrated spheres as a function of penetration length $\delta r$ (panel a) and as a function of radius of one of the spheres $R_2$ (panel b). The fit  with the use of eigenvalues of the coupled oscillators Hamiltonian (see below) is also depicted.

The crucial parameter for the developed theory is the intersection volume of two particles in the dimer, which is quadratic in penetration length $\delta r$  under the assumption that $\delta r \ll R_{1,2}$:

\be
  \label{eq_V12}
  V_{12}(R_1,R_2,\delta r) \approx \frac{\pi \delta r^2 R_1 R_2}{R_1+R_2}+ O(\delta r^3).
\ee

\subsection{How to construct the coupled oscillators Hamiltonian}

Basing on the results presented above, now we formulate the following rules  allowing to construct the ``Hamiltonian'' of the coupled oscillators model $H_{pq}$ capable to reproduce the optical phonon modes in contacting particles. The indices $p$ and $q$ in $H$ span all the modes of interest in all the particles  considered (it may exist more than two of them), e.g., $p\rightarrow(i,m)$  stands for the $m$-th mode in the $i$-th particle.

\begin{itemize}
    \item For each pair of particles $i$ and $j$ one calculates the intersection (overlap) volume $V_{ij}$. Either its approximate value given by Eq. \eqref{eq_V12} or precise value could be taken.
    
    \item For each mode $m$ of interest in $i$-th particle one writes in the diagonal element  $H_{pp}$ of the Hamiltonian [where $p=(i,m)$] its bare eigenvalue $q^2_{im}$. For optical phonons it should be taken as  the value of the size-quantization induced red shift of the given mode frequency in the $i$-th nanoparticle with respect to its bulk crystal value $\omega_0$ measured in units of cm$^{-1}$, see Eq. \eqref{eq_dispersion2}.
    
    \item For each mode $m$ one calculates the on-site volume correction (diagonal) by summing over the neighboring particles $j$: $\Delta_{p} \equiv \Delta_{im} = q^2_{im} \alpha_m \sum_j f(V_{ij}/V_i) $; after that, one adds the result to the diagonal matrix element. The factor $\alpha_m$ (see below) depends on the symmetry of the mode: s, p, etc.; $f(x)=x-(x/0.425)^2$, where the second term is an empirical correction important for relatively large penetrations.
    
    \item  For each pair of modes from different particles, one calculates the off-diagonal coupling terms $H_{pq} = - C_{pq} = - \sqrt{ q^2_{im} q^2_{jn}} \beta_{mn} V_{ij}/ \sqrt{V_i V_j} $. Here $\beta_{mn}$ are again the coefficients, dependent on the symmetry of the modes (see below).
    
    \item The eigenvalues of $H$ should be subtracted from $\omega_0$ in order to obtain the physical frequencies visible in the Raman spectra. The eigenfunctions  of $H$ reveal the amplitudes of phonon modes located at  the particles under consideration.
\end{itemize}

\subsection{Examples of construction}

According to this construction algorithm, the dimension of $H$ is the number of all modes in all particles. If we are interested, say, only in the lowest s modes, it reduces to the number of particles $N$. If one s and three p modes at each particle are considered, then the dimension of $H$ will be $4N$, etc. The shape dependent coefficients are $\alpha_s=1, \alpha_p=3$ and $\beta_{ss}=1, \beta_{sp}=3, \beta_{pp} = 4.5$.

For the simplest case of two particles and accounting for s modes only (thus the mode indices $m,n$ could be omitted), the coupled oscillators Hamiltonian reads as follows:

\be
  H = \left(
  \begin{array}{cc}
   q_1^2 - \Delta_{1} & - C_{12} \\
   - C_{12} & q_2^2 - \Delta_{2} \\
  \end{array}
\right).
\label{eq_coupledOsc_ss}
\ee

In the case of identical particles and small penetrations [$f(x) = x$], the on-site corrections and couplings are equal to each other $\Delta_{1} = \Delta_{2} = C_{12}$ and the Hamiltonian essentially coincides with the eigenvalue problem for two coupled oscillators given by Eq. \eqref{eq_newtonian_harmonic}. The only difference is negative sign in front of $k_{\rm int}$ on the diagonal. As a result, the resonant case eigenvalue that corresponds to the anti-symmetric eigenfunction does not change  (cf. Fig. \ref{fig_WFplots}) whereas for real coupled harmonic oscillators [Eq. \eqref{eq_newtonian_harmonic}] it holds for the symmetric eigenfunction. The second eigenvalue, that is lower in magnitude, corresponds to the symmetric wave function, while for real coupled harmonic oscillators, this eigenvalue with changed (higher) magnitude is related to the antisymmetric wave function. 

An additional difference appears in the case of unequal spheres. The on-site volume correction terms depend only on the intersection volume, the sphere volume and the bare eigenvalue of this sphere. Physically, it means that an additional volume in the neighboring particle becomes accessible for the phonon mode from the first nanoparticle. This volume does not depend on the precise shape of the second overlapped manifold. The coupling term is in fact the \textit{geometrical mean} of the  on-site volume corrections [within the assumption $f(x)=x$], which is equal to the on-site volume correction only for equal spheres.

For more complicated case of 2 particles with 2 modes ($m,n=1$ for s and $m,n=2$ for p$_x$ mode) at each particle, one has the Hamiltonian 

\be
  H=\left(
  \begin{array}{cccc}
    q_{1}^2 - \Delta_{1} & 0 & -C_{13} & -C_{14} \\
    0 &  q_{2}^2 - \Delta_{2} & -C_{23} & -C_{24} \\
   - C_{13} & -C_{23} &  q_{3}^2 - \Delta_{3} & 0 \\
    -C_{14} & -C_{24} & 0 &  q_{4}^2 - \Delta_{4} \\
  \end{array}
\right),
\ee
where the notation $p\rightarrow(i,m)$ has the explicit form $1\rightarrow (1,1)$, $2\rightarrow (1,2)$, $3\rightarrow (2,1)$, and $4\rightarrow (2,2)$. Among the p-modes, only those aligned along $x$ axis are considered because p$_y$ and p$_z$ modes have vanishing overlaps at small penetrations.

\begin{figure}
  \centering
  \includegraphics[width=8cm]{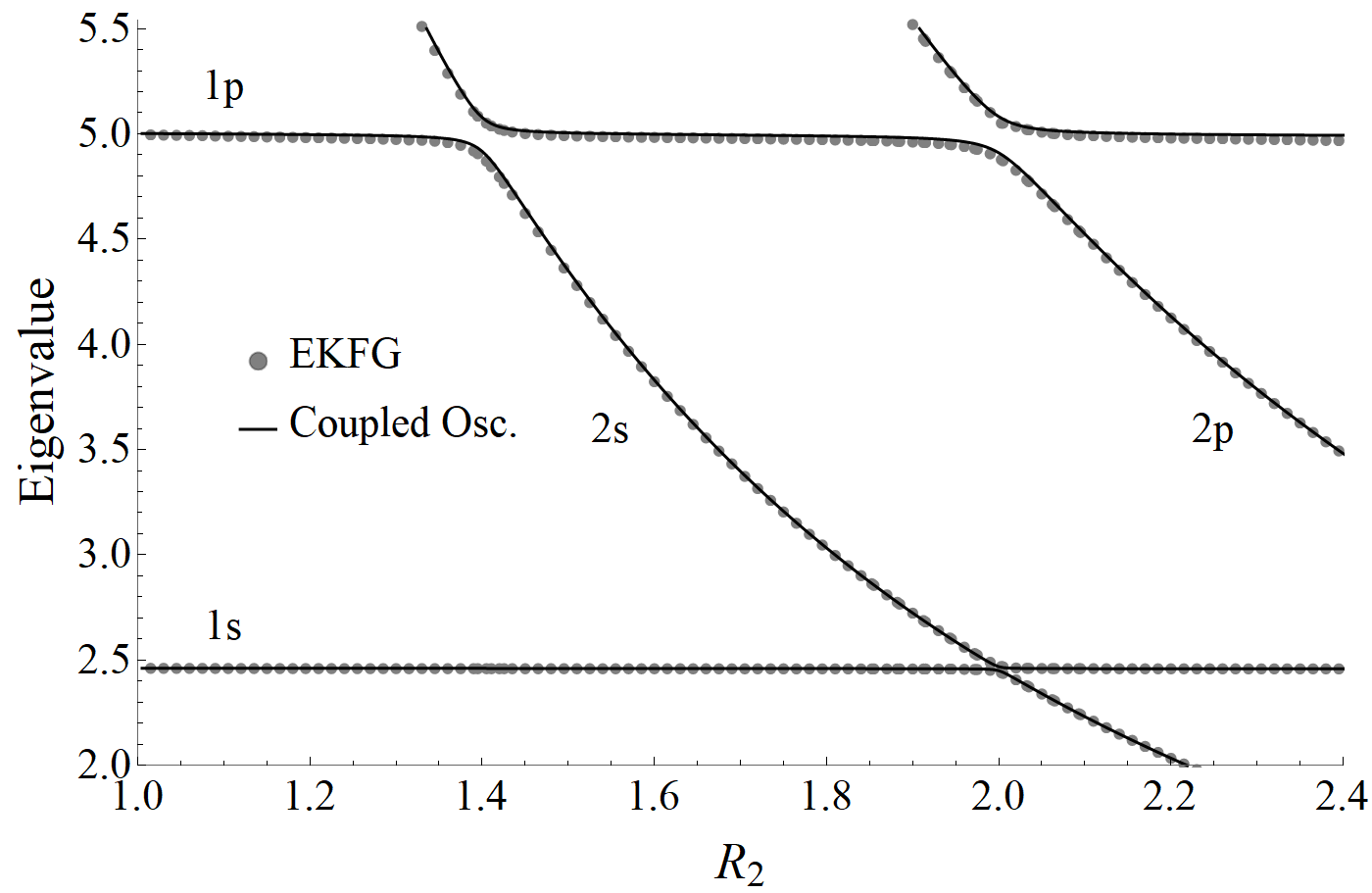}\\
  \caption{Four highest eigenvalues for nanoparticle dimer when varying the size of second particle keeping constant the size of the first one and the penetration length  ($R_1 = 2$ and $\delta r_{12}=0.2$). The dots stand for the solution of the full Dirichlet eigenvalue problem $\Delta \psi + q^2 \psi=0$. The curves give the results of the coupled oscillators model.}\label{fig_spsp}
\end{figure}

Fig. \ref{fig_spsp} shows the result of its diagonalization together with the results of EKFG numerical calculation of the Laplace eigenvalue problem. Once again, one can see a good agreement between the exact solution and the coupled oscillators approach.

Finally, we check the applicability of our approach to an important case of many particles in contact. We consider only s-modes in four identical particles located at the vertices of slightly deformed square, which form a {\it tetramer}. The corresponding Hamiltonian for small penetrations [$f(x) = x$] has the following form:

\begin{widetext}
\be
  H = \left(
  \begin{array}{cccc}
    q^2-C_{12}-C_{13} & -C_{12} & -C_{13} & 0 \\
    -C_{12} &  q^2-C_{12}-C_{24} & 0 & -C_{24} \\
    -C_{13} & 0 &  q^2-C_{13}-C_{34} & -C_{34} \\
    0 & -C_{24} & -C_{34} &  q^2-C_{24}-C_{34} \\
  \end{array}
\right).
\label{eq_Htetramer}
\ee
\end{widetext}
The on-site volume correction terms $\Delta_i$ are here written explicitly via the couplings $C_{ij}=q^2 V_{ij}/V$ to underscore the contributions of two neighbors for each particle. The mode indices are omitted similar to Eq. \eqref{eq_coupledOsc_ss}.

\begin{figure}
  \centering
  \includegraphics[width=8cm]{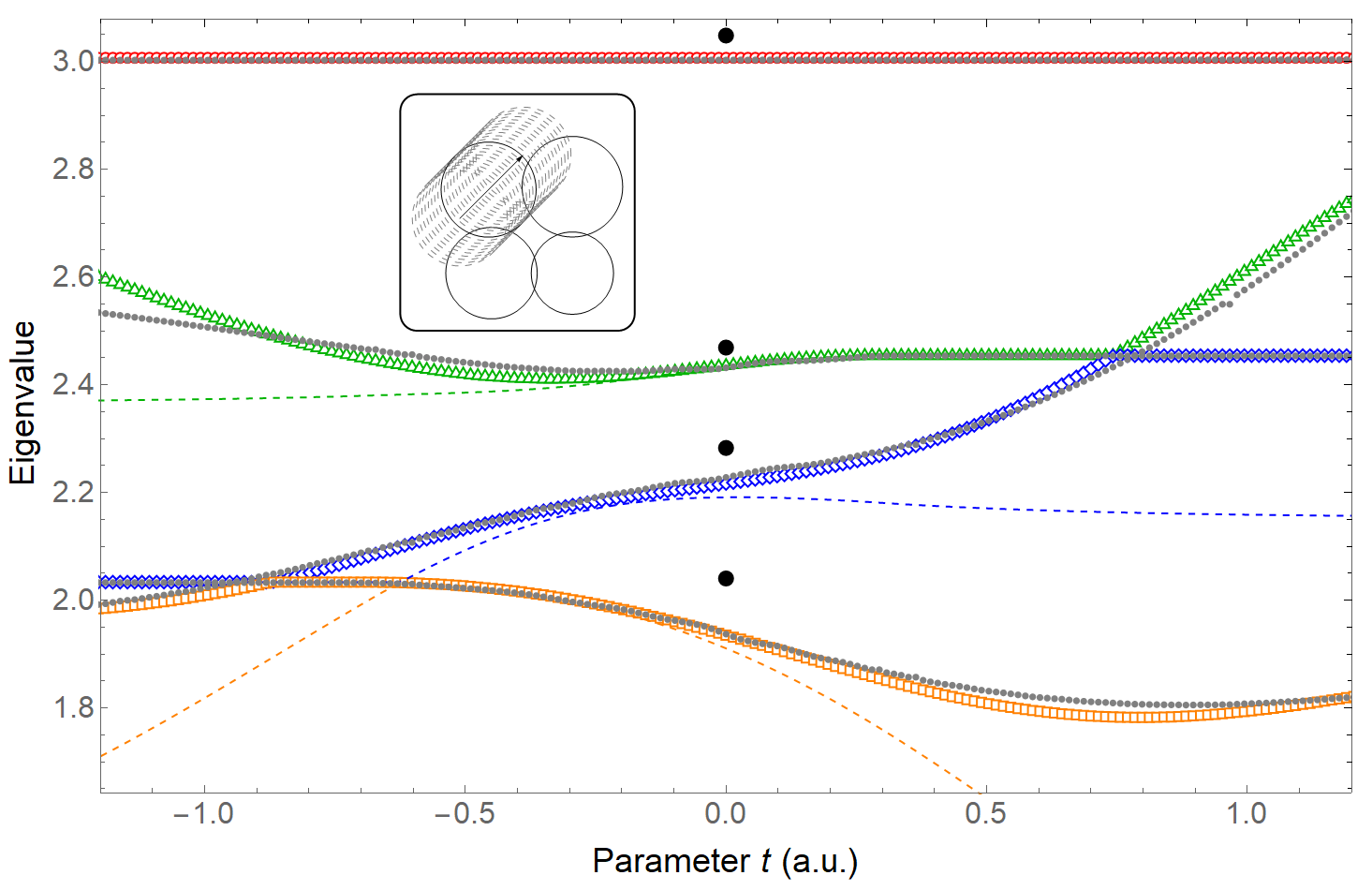}\\
  \caption{Four lowest eigenvalues for various configurations of a tetramer. Initially four particles with radii 2, 1.8, 2.2 and 2.08 are placed in  corners of rectangle in XY-plane. The sides of the rectangle could be calculated from the penetration lengths $\delta r_{12} = 0.26$ and $\delta r_{23} = 0.22$. Then, the fourth particle is moved by $t$ along horizontal and vertical axis, the trajectory of the center is depicted by an arrow in the inset giving the top view. The gray dots represent the results for the original Dirichlet eigenvalue problem $\Delta \psi + q^2 \psi = 0$. The colored dashed curves are for the 4x4 $H$ for s modes with $f(x) = x$ and approximate intersection volumes $V_{ij}$ calculation via Eq. \eqref{eq_V12}. The colored open markers are for the 4x4 $H$ for s modes with $f(x) = x - (x/0.425)^2$ and precise calculation of the intersection volumes. }\label{fig_tParam}
\end{figure}

In order to investigate even more complicated case, we study numerically unequal spheres with the Hamiltonian \eqref{eq_Htetramer} modified according to the rules formulated in previous subsection. Fig. \ref{fig_tParam} shows the result of diagonalization for the Hamiltonian constructed within the coupled oscillators model and the results of numerical solution of the eigenvalue problem for the Laplace operator with  Dirichlet boundary conditions. One sees a very good correspondence between the approaches. Only at very high penetration lengths the deviations become significant.

Unfortunately, it seems to us impossible to construct the effective perturbation theory in a spirit of Tight Binding Model \cite{ashcroft1976solid} for the media that consists of the intersected spheres, because the Dirichlet boundary conditions mimic the infinite potential barrier which leads to the divergence of overlap integrals. Even if we write the Janes-Cummings--like Hamiltonian just as a fitting model \cite{kavokin2017microcavities}

\be
  H = \left(
  \begin{array}{cc}
   k_1 & k_{\rm int} \\
  k_{\rm int} & k_2 \\
  \end{array}
\right),
\label{eq_JCH}
\ee
then the two levels will repel each other symmetrically in resonance case, cf. Eqs. \eqref{eq_newtonian_harmonic} and \eqref{eq_inresonance} and  Fig. \ref{fig_modelcase}(a), where our analysis shows that one level stays nearly intact upon the coupling value increase.

\section{Dynamical Matrix Method}

The more direct atomistic DMM method of optical phonons treatment is even better adapted for precise calculations of vibrational modes in nanoprticles including the intermediate wavelengths regime and phonon polarizations. It also allows to consider the loose arrays of faceted nanoparticles weakly interacting via the Van der Waals forces.

\subsection{Method formulation}

Dynamical Matrix Method (DMM) is successful in obtaining the vibrational modes of molecules, atomic clusters and nanoparticles. It is based on writing the second Newton's laws for all atoms with the forces caused by bonds stretching and bending of valence angles. Incorporating into calculations the conventional Keating model \cite{keating1966effect,martin1970elastic,anastassakis1990piezo,kane1985phonon} allows us to write  these equations of motion  in the form:

\begin{equation}\label{eq_DMM}
m  \, \Ddot{r}_{p} = M_{pq}r_{q},
\end{equation}
with the designations $p = (i, \alpha)$ and $q = (j, \beta)$, where Latin letters enumerate the atoms and Greek letters span over the Cartesian coordinates: $\alpha,\beta = x,y,z$. The dynamical matrix is given by

 \begin{equation}\label{DMM}
M_{pq} = \sum_{j=1}^{N} \sum_{\beta =x,y,z} \frac{\partial^2 \Phi({{\bf r}_1,{\bf r}_2},...)}{ \partial r_{j, \alpha} \partial \, r_{j, \beta}} r_{j, \beta}.
\end{equation}
In the equations above $m$ is the mass assumed to be the same for all atoms, $\Phi$ is total potential energy of nanocrystallite expressed via displacements ${\bf r}_i$ of  atoms from their equilibrium positions. More specifically $\Phi$ is a sum of pairwise interaction energies for bond stretching and depends on the positions of three atoms for valence angles deformations.

Equation \eqref{eq_DMM} can be solved for tracing the time dynamics from some initial conditions or, in the frequency representation, for obtaining the phonon frequencies and eigenmodes:
 \begin{equation}\label{eq_DMM_eigen}
m \omega^2 r_{p} = - M_{pq}r_{q}.
\end{equation}
Vector $r_p$ contains all information about phases and directions of atomic displacements.

\subsection{Results for cojoined spheres. Comparison with COM}

It is natural to verify the proposed coupled oscillators model by the comparison of its predictions with the numerical results of atomistic DMM. When constructing the coupled oscillators Hamiltonian using Eq. \eqref{eq_coupledOsc_ss}, one can directly take the Dirichlet problem bare eigenvalue $q^2$ for separate particle as its energy red shift \mbox{$q^2 = \omega_0 - \omega$} with respect to the optical phonon frequency in the BZ center $\omega_0$ [see Eq.~\eqref{eq_dispersion2}; here we put $\alpha=1$ and measure $q^2$ in cm$^{-1}$]. Note, that below we shall discuss the results in terms of the optical phonon frequencies taking the diamond  with $\omega_0 = 1333$~cm$^{-1}$ as an example. We choose the Keating model parameters as in Ref.~\cite{ourDMM}.

\begin{figure}[t]
  \centering
    \includegraphics[width=8cm]{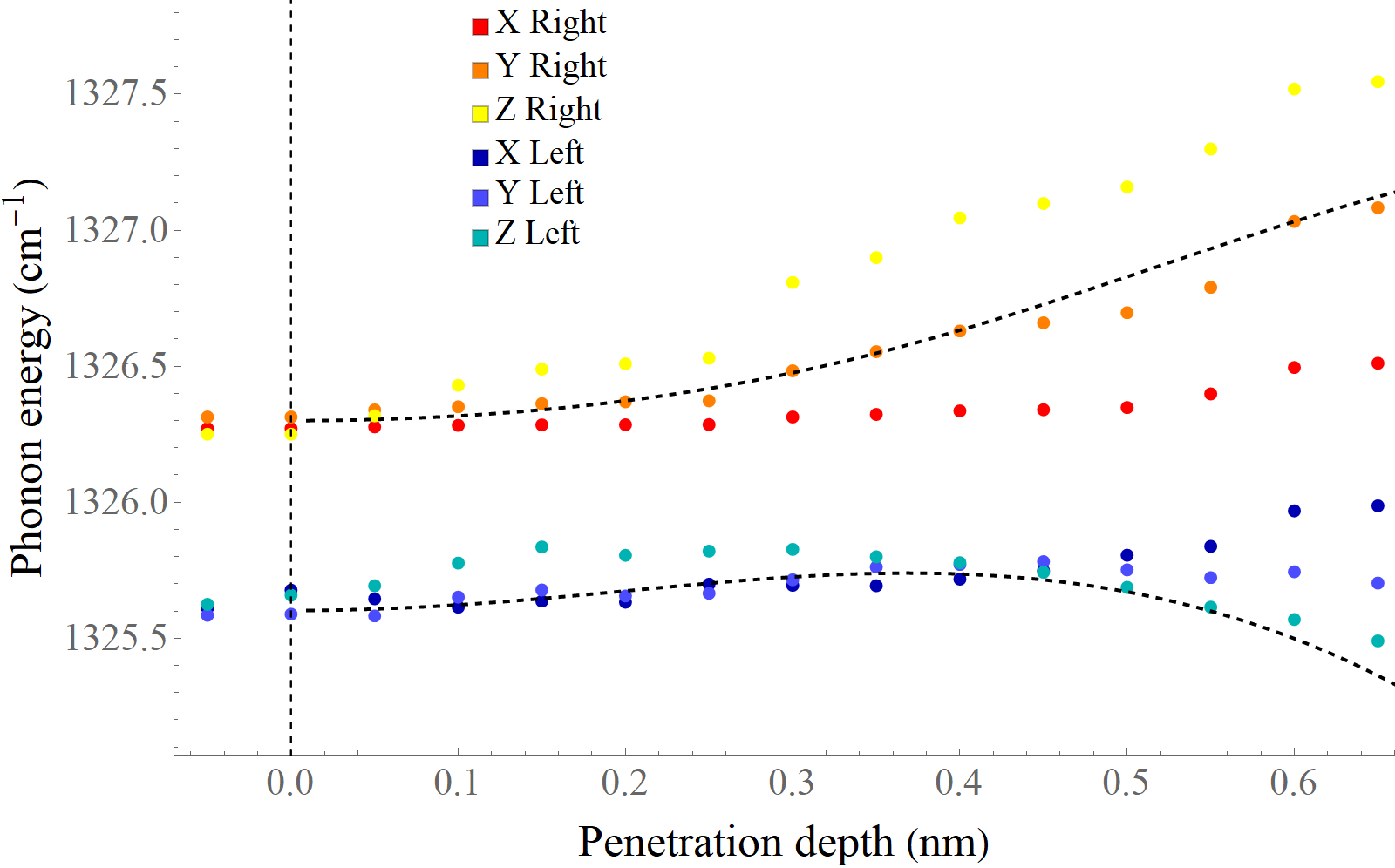}
  \caption{Six highest phonon frequencies of 2.4~nm and 2.3~nm spherical diamond particles in contact as a function of penetration depth $\delta r$ (here $\delta r < 0$ corresponds to isolated particles and $\delta r=0$ to contact via single atom). Different colors correspond to various phonon polarizations. Red, orange and yellow colors are for eigenfunctions mainly localized inside the 2.4 nm particle (right particle). Blue tones correspond to the 2.3 nm particle (left particle). The dashed curve is obtained based on the coupled oscillators model.}
\label{fig_DMM_on_dx_analytCU}
\end{figure}

Fig.~\ref{fig_DMM_on_dx_analytCU} shows six highest eigenvalues for a couple of diamond particles as a function of penetration length $\delta r$ obtained using the DMM approach (the lowest eigenvalues $q^2$ of the Dirichlet problem in terms of EKFG approach due to the dispersion law with negative effective mass). The obtained data is compared with the predictions of COM. One sees that for various optical phonon polarizations the couplings have differing magnitudes. Still the COM works semi-quantitatively. The out-of-resonance case similar to what we expect from   Eq.~\eqref{eq_out_of_resonance} is visible at small penetrations where both energies are shifted equally and the shift is proportional to $\delta r^2$.

\begin{figure}[t]
  \centering
    \includegraphics[width=8cm]{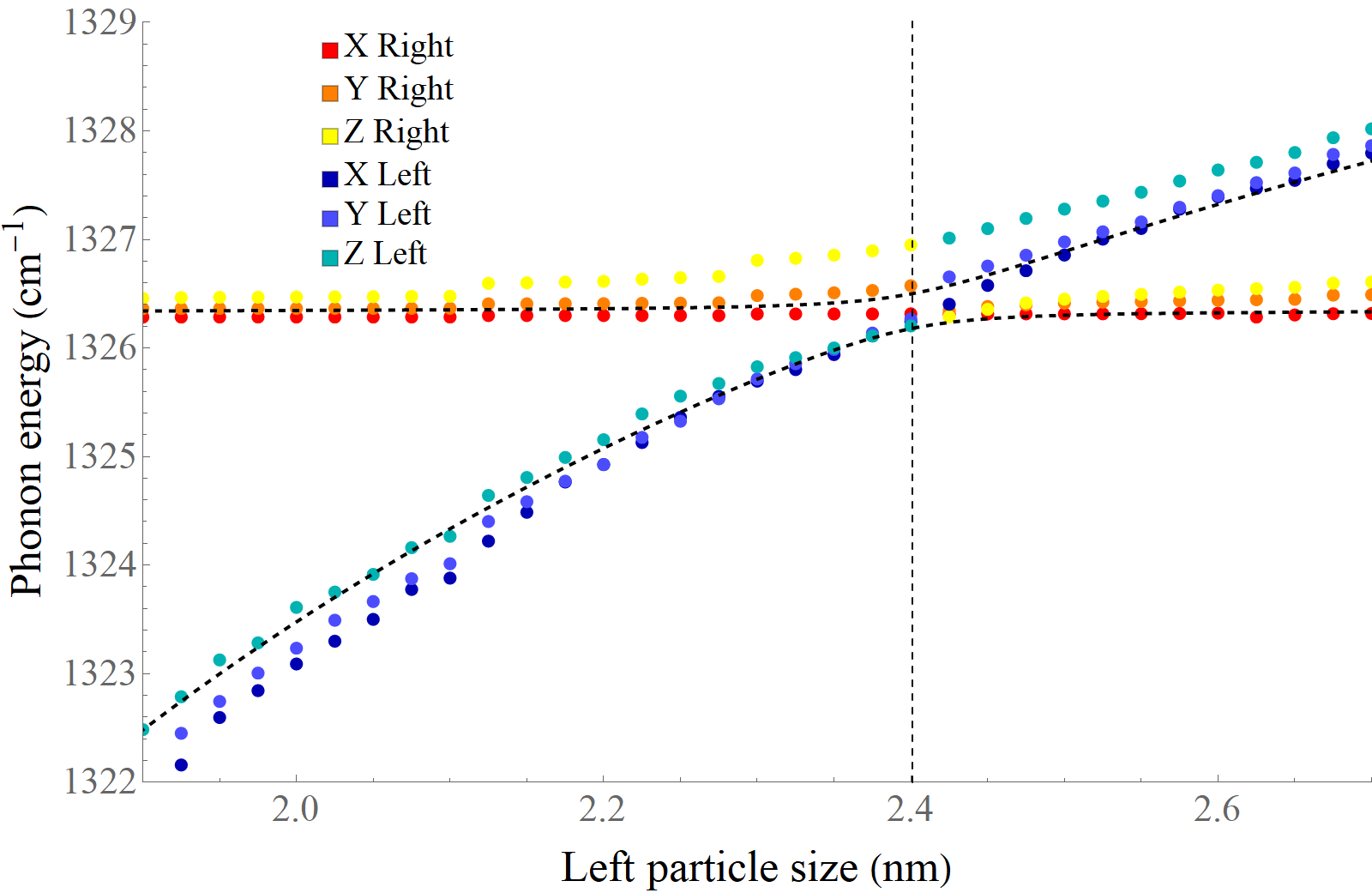}
  \caption{Six highest phonon frequencies of the nanoparticle dimer as a function of the diameter of the second (left) particle. The size of the right particle is 2.4 nm. The dashed curves are obtained based on the coupled oscillators model.}
\label{FigS}
\end{figure}

One sees that polarization results in more complicated behavior of modes hybridization. COM describes the situation for all separate modes qualitatively and for their sum quantitatively. Fig.~\ref{FigS} illustrates the same idea, investigating the case when we vary the size of one of the particles whilst the penetration length remains intact.

\subsection{Results for weakly interacting faceted particles}

Another important and physically meaningful case is attaching of {\it faceted particles}. The underlying mechanism of their interaction can be the Van der Waals forces or the covalent ones similar to the dimer case. Below we shall use $k$ for the rigidities of interparticle bonds and $K$ for the regular intraparticle (covalent) ones.

Fig.~\ref{fig_DMMcubes_shiftskK} shows the frequency shifts of three highest phonon modes in each cubic particle of a touching couple. The sizes of particles are 1.8 nm and 1.7 nm, respectively. For $k/K\ll1$ the frequencies are all blue shifted linearly with {the bond strength $k$, which qualitatively corresponds to the out-of-resonance case. Physically, the van der Waals interaction constant $k$ is about three orders of magnitude lower than the covalent one \cite{anastassakis1990piezo,mayo1990dreiding}.

One can estimate the coupling-induced frequency shift as \mbox{$\delta \omega \propto \omega_0 \psi_{\rm surf}^2 V_{\rm inter}\cdot(k/K)$}, where $\psi_{\rm surf}$ is an estimation of the atomic displacements magnitude at the surface and $V=S\cdot a_0$ is the effective interaction volume ($S$ is contact surface and $a_0$ is the lattice parameter):

\be
  \delta \omega = {\rm const} \cdot \omega_0 \left( \frac{1}{\sqrt{L^3}} \cdot \frac{a_0}{L} \right)^2 \cdot (S \cdot a_0) \cdot \frac{k}{K}.
\ee
For the case of faceted particles whose facet surface (and therefore) the contact area is proportional to the size $L$, the following scaling takes place:
\be
  \delta \omega = {\rm const} \cdot \omega_0 s_{\rm rel} \left( \frac{a_0}{L} \right)^3 \cdot \frac{k}{K}, 
\ee
where $s_{\rm rel} = s/S$ is the percentage of surface experiencing a contact with the neighboring particle.

\begin{figure}[t]
  \centering
  \includegraphics[width=7cm]{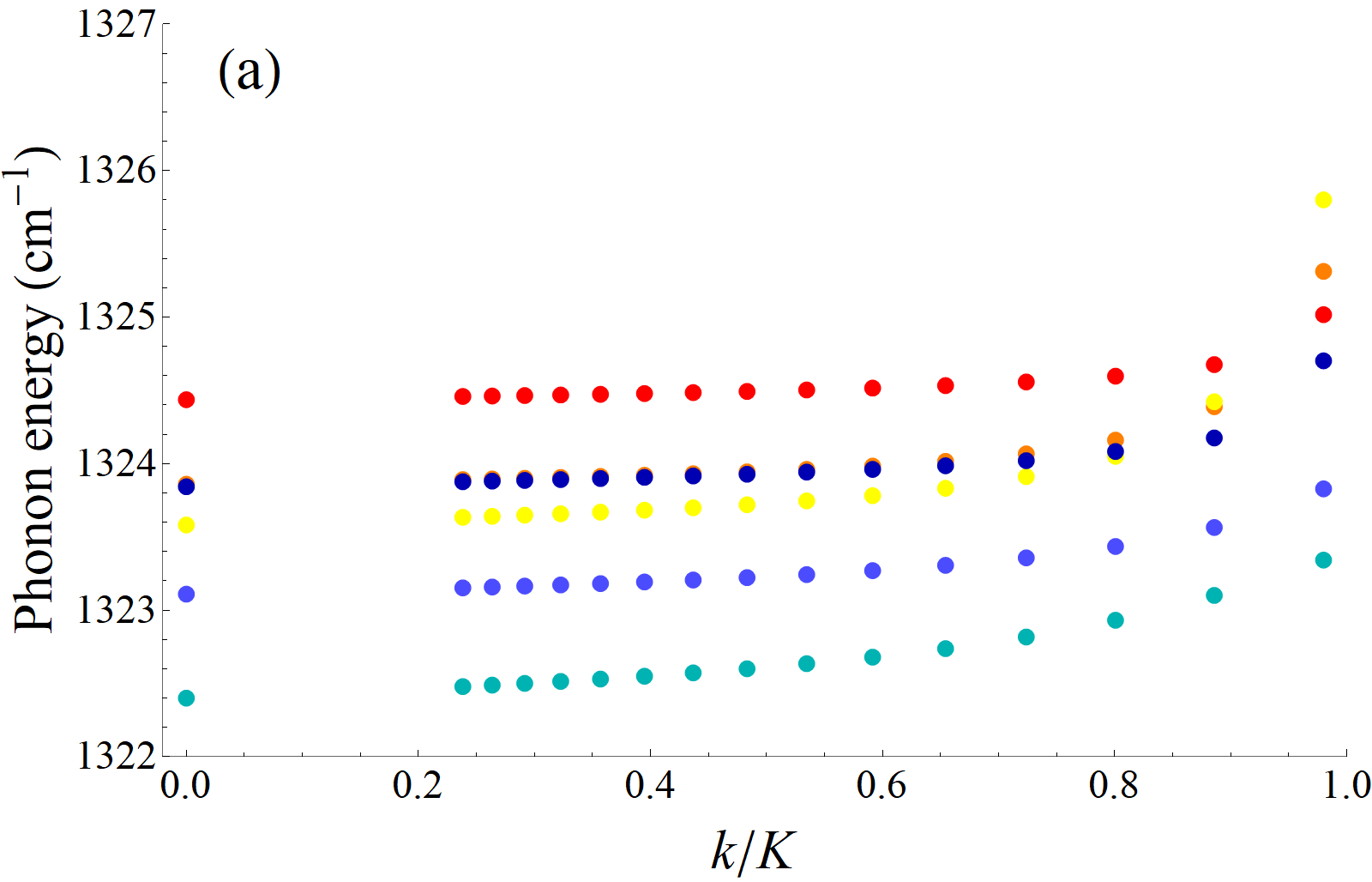}\\
  \includegraphics[width=7cm]{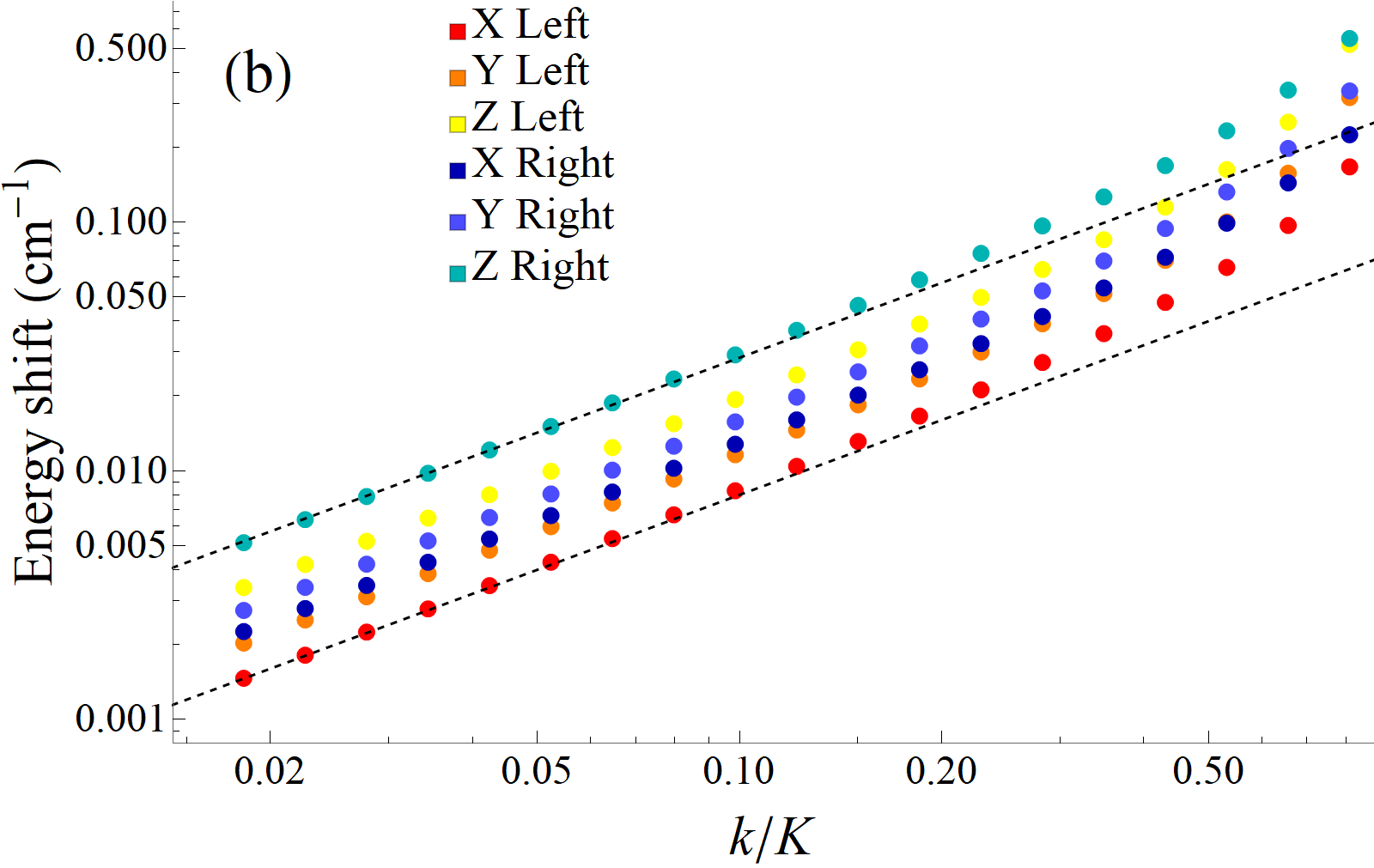}
  \caption{(a) Energies of six highest phonon modes for two weakly interacting cubic particles of the sizes 1.9 and 1.8 nm as functions of $k/K$ ratio. (b) The same as (a) but for energy shifts with respect to $k=0$ case (noninteracting particles). Eye-guides for linear dependencies are given.}
\label{fig_DMMcubes_shiftskK}
\end{figure}

\begin{figure}[t]
  \centering
  \includegraphics[width=7.cm]{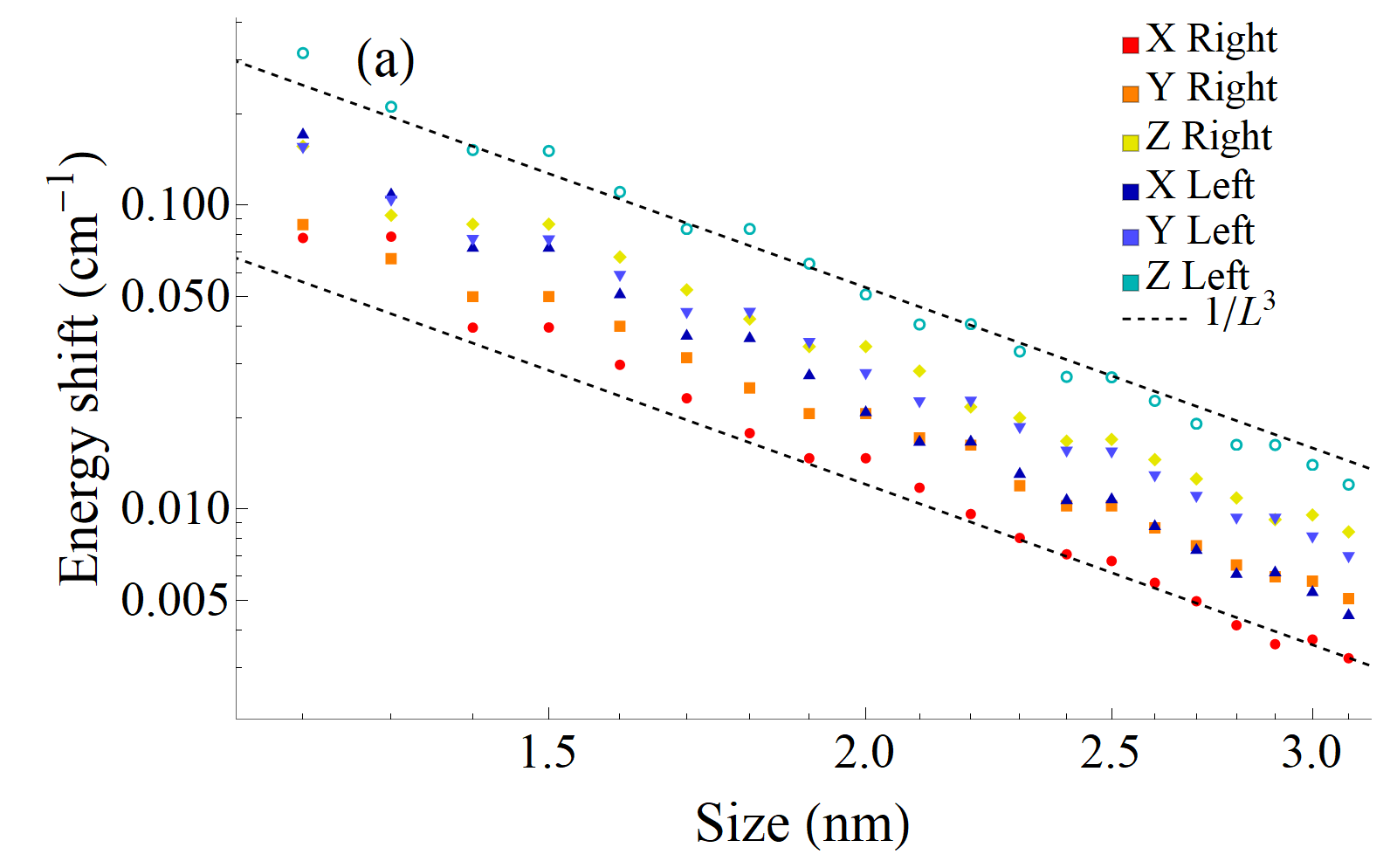} \\
  \includegraphics[width=7.cm]{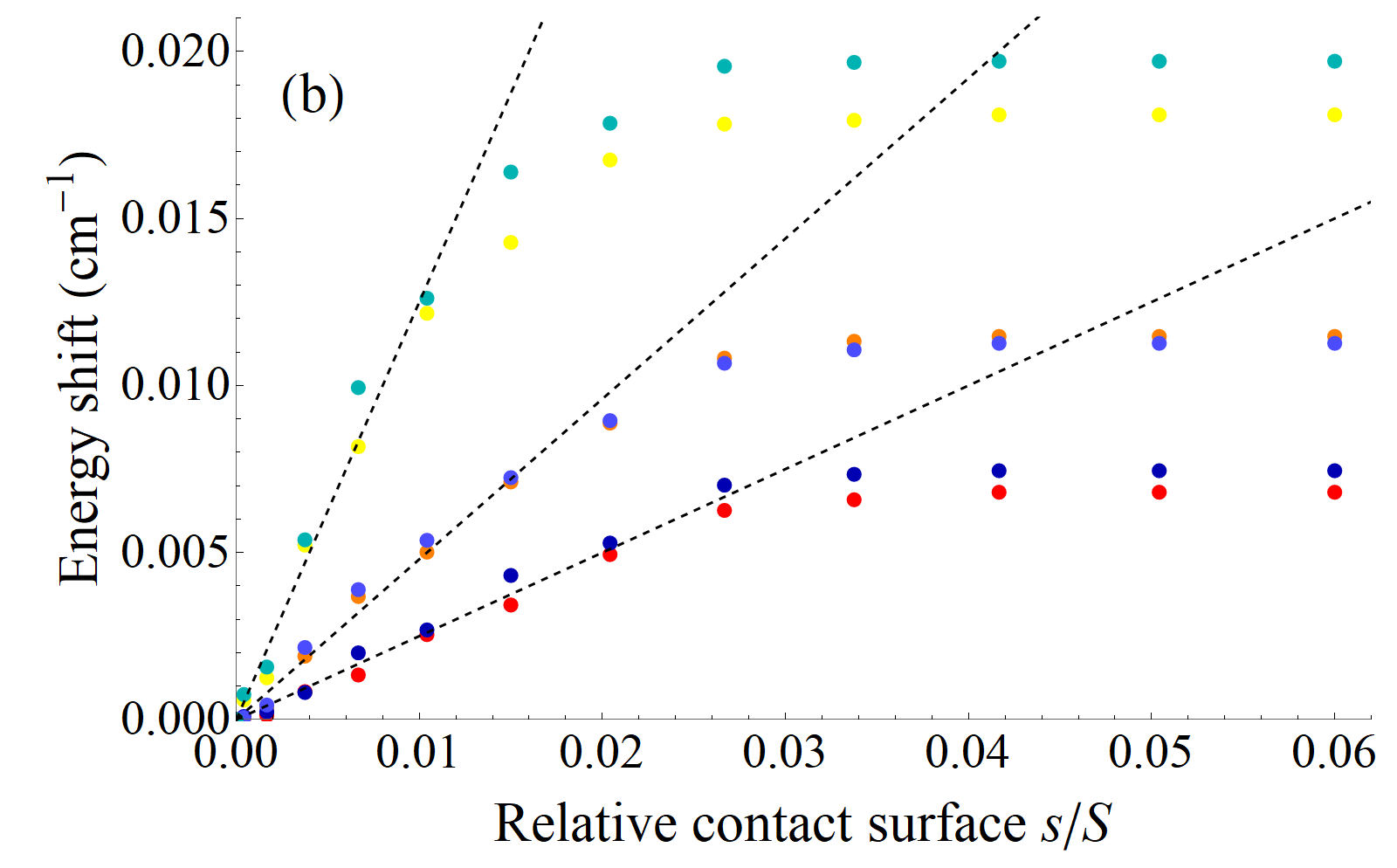}
  \caption{(a) The energy shifts of 6 highest phonon modes for two cubic particles of size $L$ (given by the horizontal axis) and $0.9L$ in contact, $k/K$ ratio is 0.2. Shifts are calculated with respect to the non-interacting particles $k/K=0$. The dashed lines give the eye-guides for $L^{-3}$ power law. (b) The eigenvalue shift as a function of contact surface $s$ with respect to the full surface $s/S=l^2/(6L^2)$. Two equal 3 nm cubes are contacting only in the central square-shaped region (of size $l<L$) of their joint facet, the bond strengths in contact are $k/K=0.333$. For large range of parameters the energy shift is linear in $s/S$, the corresponding eye-guides are depicted.}
\label{fig_DMMcubes_investigation}
\end{figure}

In Fig.~\ref{fig_DMMcubes_investigation} one can see the shifts saturation occurring  because the phonon wave function (the standing wave in the shape of product of three cosine functions for cubes) decreases from the facet center to its edges. As a result, the overlap of wave functions of two particles is defined by the central regions of the facets. In practice, the effects should be even smaller because of a mismatch between contacting facets, their corrugations and presence of functional groups. We conclude that without a big spot of covalent bonds, the effect of contacts on optical phonons in general and on Raman spectra in particular is negligible taking into account the present accuracy of measurements. From a point of view of the possibility to build up an adequate theory of modes propagating in such a media, it implies that one can use for this sake the unperturbed eigenfunctions and eigenvalues of isolated particles.

\section{Dimer Raman spectra}

In this section we study the effect of nanoparticles dimerzation on the Raman spectra. Silicon nanoparticles with the mean size $4$~nm are normally distributed around this value (FWHM 5\% of mean value). The corresponding Raman spectra~\cite{xia1995phonon,paillard1999improved,faraci2006modified,gupta2009modified,volodin2013improved} are red shifted by approx. 5~cm$^{-1}$ with respect to the bulk silicon with the peak centered at 520~cm$^{-1}$. The nanoparticle dimers with penetration length distributed uniformly in the range from 0 to 1~nm is compared with the abovementioned case. Within EKFG, the spectra are calculated using the standard procedure~\cite{ourEKFG} formulated as follows. The intensity of each mode is given by a square of the wave function volume integral $|\int \psi ({\bf r}) d{\bf r} |^2$ (cf. with the structure factor in the conventional scattering problems for $q=0$) and then the summation over all modes is performed in order to incorporate the phonon peak intensities and their positions into the Raman peak. The Raman spectrum can be also obtained using COM. Within the isotropic s-mode approximation resulting in the 2x2 Hamiltonian for the dimer, the intensity of each of two modes is $\propto |\psi_1 V_1 + \psi_2 V_2|^2$, where $\psi_i$ stands for the components of  the ``spinor'' $\psi$. Fig. \ref{fig_ramanSi} shows the Raman spectra of free particles and nanoparticle dimers.

\begin{figure}[t]
  \centering
  \includegraphics[width=0.48\textwidth]{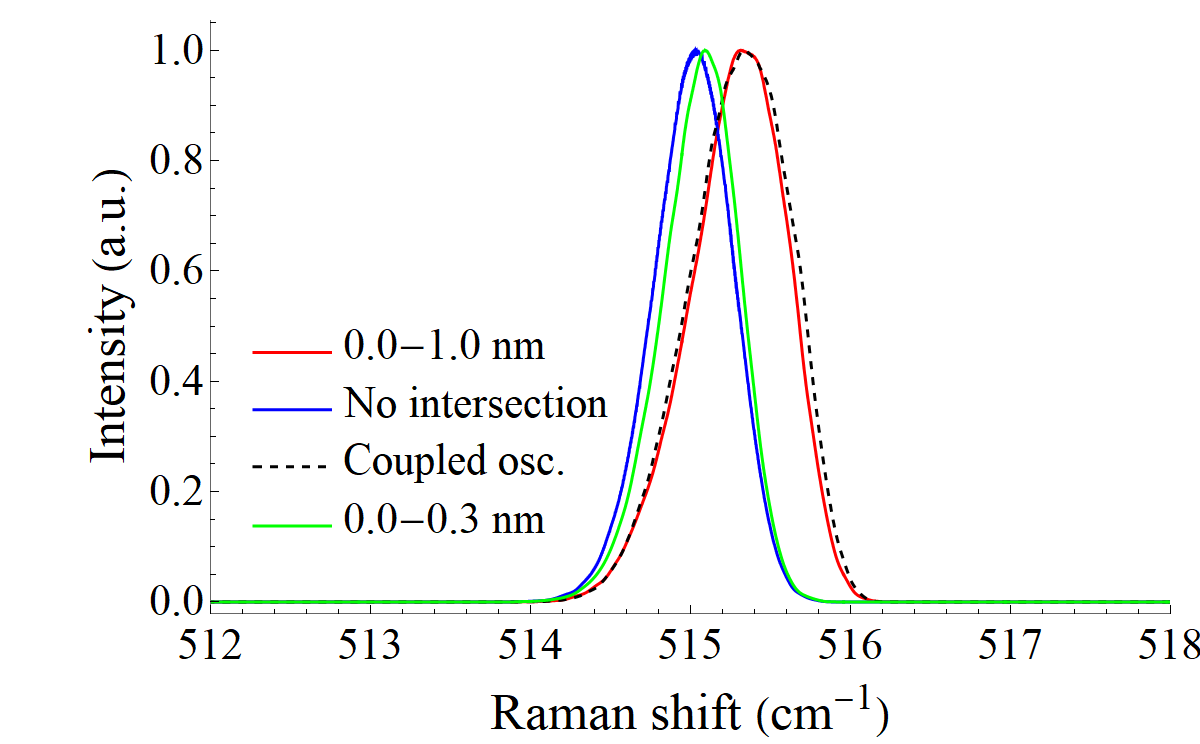}
  \caption{Raman spectra of $4\pm0.1$~nm silicon particles with and without (blue curve) dimerization. Red curve gives the EKFG approach yield for penetration lengths varying randomly from 0 to 1~nm. Dashed black curve is for the Raman spectrum obtained using the coupled oscillators approach with the same size and penetration length distributions. Green curve is for smaller penetration lengths: from 0 to 0.3~nm.}
\label{fig_ramanSi}
\end{figure}

One sees the full agreement between EKFG and COM. Dimerization results in decreasing of the red shift of the peak and in changing of its shape. Effectively, it can be explained by a higher volume accessible for the phonons which suppresses the confinement and reduces the size quantization effects. The effect of dimerization is at the level of experimental accuracy of modern spectrometers (0.3 cm$^{-1}$), however the peaks differ {\it significantly} in their shape. The peak asymmetry coefficient for dimers is 0.37 when for free particles it is 0.22, which in our opinion rules out the possibility of confusion. We believe that incorporation into the theory the ``out of shell'' effects~\cite{our3} making the phonon lines even more asymmetric will further improve the situation.

\section{Discussion and Conclusions}

In the present paper, we have studied the behaviour of optical phonons in contacting nanoparticles. 
Such phonons obey the Euclidean metric Klein-Fock-Gordon equation leading to the Laplace eigenvalue problem \mbox{$\Delta \psi + q^2 \psi = 0$} with Dirichlet boundary conditions $\psi|_{\partial \Omega}=0$.We have used this approach as a frame and reference point comparing its numerical solution for interpenetrating spheres with the results of phenomenological but technically simpler method, which we called the coupled oscillators model (COM). We have formulated the phenomenological COM where each mode in each particle is brought into correspondence with an oscillator of a given frequency. The particle contacts cause appearance of additional oscillator couplings proportional to the volume of particle overlaps. For not too large overlaps, the formulated algorithm describes pretty well the behavior of levels' eigenvalues (\textit{i.e.,} s and p symmetric vibrational modes) as a function of particles sizes and penetration depth, being in a good agreement with the original Dirichlet problem. In the case of many particles, the formulated rules do not loss their accuracy and thus the coupled oscillators model can be used for large arrays of contacting nanoparticles. The COM approach works well for s and p levels up to ratio of penetration depth to particle size $\delta r/R \lesssim 0.15$. At higher ratios, the complex geometry of p modes starts playing its role, and more sophisticated functions of intersection volumes are required to construct the COM Hamiltonian. However, the bands of s and p modes remain well separated for not too big scatter in size (when the levels of differing symmetry do not overlap without interaction). For s levels only and pointed out condition on scatter in size,  COM works well up to $\delta r/R<0.5$.

To solidify our findings, we have compared the results of COM with the yield of microscopic Dynamical Matrix Method for lattice vibrations. Also, DMM has been used to consider the special but physically relevant case of faceted particles with weak van der Waals interaction. Only small frequency shifts lying far beyond the accuracy of the Raman spectroscopy have been obtained in this case. We conclude that the Raman spectra of nanopowders can be interpreted neglecting the effects of particle-particle contacts. On the contrary, for QD Nanocrystal Solids and porous materials (the latter can be considered as a network of cojoined nanoparticles) the effects of the optical phonons hybridization should be strong enough to result in phonon long-distance propagation accompanied by the Raman peak shift and broadening with respect to the powder of nanoparticles of comparable size.

It also follows from our analysis that the straightforward formulation of the perturbation theory in a spirit of quantum mechanics-like Tight Binding Model for diatomic molecule is not possible because the Dirichlet boundary conditions correspond to infinite potential barriers. Concerning the general properties of Laplace operator eigenvalue problem, in literature there exist only general statements like Rayleigh–Faber–Krahn inequality~\cite{henrot2006extremum}: the ball of the required dimension has the lowest eigenvalue when the body volume is fixed. The problems of minimization and maximization of the first Dirichlet eigenvalue for Laplacian in the body with an obstacle~\cite{henrot2017optimizing} and minimization of the eigenvalues beyond the first one~\cite{henrot2003minimization} (see also the bibliography therein) were also considered, but they do not help to quantify the behavior of eigenvalues and eigenfunctions of intersected spheres when varying their size mismatch and penetration parameter. Finally, the physical problem of the capacitance of joined spheres \cite{felderhof1999electrostatic} also deserves here to be mentioned, but is also does not provide the required asymptotic behavior of eigenvalues for the interpenetrating spheres.

As far as the theory describes the behavior of \mbox{$\Delta \psi + q^2 \psi = 0$} equation eigenvalues and eigenfunctions in the manifold of the intersected spheres, our results can be also applied to any problem leading to this mathematical physics problem. Along with the considered case of phonons in the interpenetrating nanoparticles, the example of such problem is electronic levels structure in quantum dot molecules (stationary Schr\"odinger equation). The obtained asymptotic behavior and ``Hamiltonian'' construction rules are important for the mathematical physics itself. Interestingly, the effective Hamiltonian appears to be inspired and close to elastic problem of coupled oscillators rather than to quantum mechanical tight binding-like perturbation theory.

The developed theory is applicable beyond 3D. For instance, in the two dimensional space COM-like Hamiltonian can be constructed for the polariton molecules\cite{sala2015spin} and graphene\cite{sala2015spin,nalitov2015spin,nalitov2015polariton}, which is important to account for such problems as engineering the effective gauge fields in such structures \cite{jamadi2020direct}. However, in lower dimensions due to higher fraction of the wave function exposed to the overlap area, the dependencies of matrix elements on the penetration length and overlap volume can be more complex.


\begin{acknowledgments}

The reported study was supported by the Foundation for the Advancement of Theoretical Physics and Mathematics ``BASIS''. S.V.K. conducted his research thanks to the IBS Young Scientist Fellowship (IBS-R024-Y3-2022).

\end{acknowledgments}

\bibliography{bib}
\end{document}